\documentclass{article}

\PassOptionsToPackage{numbers}{natbib}

\usepackage{amssymb,euscript,latexsym,amsmath} 
\usepackage{graphicx} 
\usepackage{epstopdf}
\usepackage{bm} 
\usepackage{enumerate}
\usepackage{color} 
\usepackage{setspace} 
\usepackage{subfigure} 
\usepackage[
pdftitle={VNLS},
pdfcreator={pdflatex},
pdfsubject={preprint},
hyperindex = {true},
colorlinks = {true},
linkcolor = {blue},
citecolor = {blue}
]{hyperref}
\usepackage{array}
\usepackage{cancel}
\usepackage{wrapfig}
\usepackage{floatrow}
\usepackage{stmaryrd}
\usepackage{bbm}

\usepackage{gensymb}
\usepackage[dvips]{epsfig}
\usepackage{epstopdf}
\usepackage{setspace}
\usepackage{subfigure}
\usepackage{lipsum}
\usepackage{wrapfig}
\usepackage{floatrow}
\usepackage[shortlabels]{enumitem}
\usepackage{algorithm, algpseudocode}

\newcommand{\ketbra}[2]{\mathinner{|{#1}\rangle\langle{#2}|}}

\DeclareMathOperator*{\var}{var}

\newcommand{\abs}[1]{\left\lvert #1 \right\rvert}
\newcommand{\norm}[1]{\left\lvert\left\lvert #1 \right\rvert\right\rvert}
\newcommand{\ip}[1]{\left\langle #1 \right\rangle}
\newcommand{\paren}[1]{\left( #1 \right)}
\newcommand{\braces}[1]{\left\{ #1 \right\}}
\newcommand{\bracket}[1]{\left[ #1 \right]}
\newcommand{\bra}[1]{\left\langle #1 \right\rvert}
\newcommand{\ket}[1]{\left\lvert #1 \right\rangle}

\usepackage[final]{neurips_2021}

\usepackage[utf8]{inputenc} 
\usepackage[T1]{fontenc}    
\usepackage{hyperref}       
\usepackage{url}            
\usepackage{booktabs}       
\usepackage{amsfonts}       
\usepackage{nicefrac}       
\usepackage{microtype}      
\usepackage{xcolor}         

\title{Toward Neural Network Simulation of Variational Quantum Algorithms}

\author{Oliver Knitter$^1$, \, James Stokes$^1$, \, Shravan Veerapaneni$^{1,2}$\\
$^1$Department of Mathematics, University of Michigan, Ann Arbor, MI 48109\\ $^2$Flatiron Institute, Simons Foundation, New York, NY 10010\\
\texttt{\{knitter, stokesjd, shravan\}@umich.edu}\\
}

\begin{document}
\pdfoutput=1

\setlength{\lineskip}{0pt}

\maketitle

\begin{abstract}
Variational quantum algorithms (VQAs) utilize a hybrid quantum--classical architecture to recast problems of high-dimensional linear algebra as ones of stochastic optimization. Despite the promise of leveraging near- to intermediate-term quantum resources to accelerate this task, the computational advantage of VQAs over wholly classical algorithms has not been firmly established. For instance, while the variational quantum eigensolver (VQE) has been developed to approximate low-lying eigenmodes of high-dimensional sparse linear operators, analogous classical optimization algorithms exist in the variational Monte Carlo (VMC) literature, utilizing neural networks in place of quantum circuits to represent quantum states. In this paper we ask if classical stochastic optimization algorithms can be constructed paralleling other VQAs, focusing on the example  of the variational quantum linear solver (VQLS). We find that such a construction can be applied to the VQLS, yielding a paradigm that could theoretically extend to other VQAs of similar form.
\end{abstract}

\section{Introduction}
 
 Recent impressive progress in quantum computational technology has led to renewed interest in quantum algorithm research. Much of this excitement stems from
 the development of quantum algorithms exhibiting exponential speedups over their classical counterparts, including algorithms for large-scale linear algebra \cite{harrow2009quantum}. Furthermore, plausible complexity-theoretic assumptions strongly suggest \cite{shepherd2009temporally} that quantum computers are capable of preparing quantum states whose output probability distributions are hard to sample classically. Despite this vast potential, quantum algorithm design suffers from a few caveats.
 
 Firstly, exponential acceleration requires {\em fault-tolerant} quantum computation, which necessitates access to a prohibitively large number of physical qubits. Even under the assumption of polylogarithmic overhead in the asymptotic limit of problem size \cite{reascended2007fault}, the necessary resources for solving linear systems of practical utility vastly exceed the likely near-term capabilities of noisy intermediate-scale quantum (NISQ) devices. Secondly, the conjectured hardness results \cite{shepherd2009temporally} for classical sampling pertain to probability distributions with no known practical utility, without any guidance toward which quantum states should be targeted for practical problems.
 
 These two obstacles motivate a new research direction called {\em variational quantum algorithms} (VQAs), wherein one encodes a computational problem as a stochastic optimization problem whose solution is an unknown quantum state; in particular, the probabilistic nature of quantum states dictates that this solution be encoded in the output probabilities of the optimal state, which may be estimated through Monte Carlo sampling. Modeling quantum states classically requires overcoming a curse of dimensionality, since the number of dimensions of a multi-qubit state space grows exponentially with the number of qubits. The potential computational advantage of VQAs stems from their reliance on a Quantum Processing Unit (QPU) to perform state preparation within a hybrid quantum--classical framework, where the CPU performs gradient-based updates in order to direct the QPU toward the optimal quantum state. Variational algorithms have already been designed with the goal of solving hard combinatorial optimization \cite{farhi2014quantum} and for ground state preparation in quantum chemistry \cite{peruzzo2014variational}; more recently, VQAs have emerged for solving fundamental linear algebra problems in high dimension, such as matrix--vector products \cite{xu2019variational}, solutions of linear systems \cite{bravo2019variational} and even singular value decomposition (SVD) \cite{bravo2020quantum, wang2020variational}. Nonetheless, the existence of a quantum computational advantage---as well as its underlying theoretical justification---has yet to be demonstrated for VQAs.
 
 To better understand this potential for quantum advantage, we utilize the close relationship between VQAs and variational quantum Monte Carlo (VQMC), an area that has shown significant recent progress in expanding its capabilities to problems beyond its traditional purview \cite{carleo2017solving}. Similarly to VQAs, the VQMC overcomes the curse of dimensionality by alternating between gradient-based optimization and Monte Carlo sampling from a parametrized quantum state. However, unlike VQAs---whose computational advantage hinges on the conjectured difficulty of sampling from a parametrized quantum circuit---the VQMC gains its power by modeling the quantum state using multi-layer parametrized neural networks \cite{carleo2017solving}. The exploitation of flexible neural networks as many-body trial wavefunctions has made it possible to leverage the enormous success of machine learning (ML) in solving a variety of quantum many-body eigenvalue problems \cite{carleo2017solving}.
 The close parallels between ideas developed in the very different fields of VQAs, VQMC, and ML has yielded many opportunities for technology transfer between the three, including the discovery of a VQMC-inspired natural gradient optimization algorithm for VQAs \cite{stokes2020quantum} and an expanding list of VQA-inspired algorithms for combinatorial optimization \cite{gomes2019classical, zhao2020natural,hibat2021variational} and partial differential equations \cite{zhao2022quantum}.

 While VQMC has traditionally focused on ground state preparation and solving the initial value problem for the time-dependent Schr\"{o}dinger equation, there is no fundamental obstruction to tackling other linear algebra problems of practical concern,
 and we strive to bridge this gap by pursuing VQMC approaches to linear algebra in exponentially high dimensions. Choosing to focus on VQMC isolates the question of quantum advantage from the inherent advantage of Monte Carlo for overcoming the curse of dimensionality. Using a recently discovered Rayleigh quotient reformulation, we introduce a VQMC-based solver for the linear system $Ax = b$, where $A$ is a large row-sparse matrix, which we call the Variational Neural Linear Solver (VNLS). 
 The VNLS provides a classical benchmark for assessing the quantum computational advantage of the Variational Quantum Linear Solver (VQLS) \cite{bravo2019variational}, which is a proposed VQA for solving sparse high-dimensional linear systems. The VNLS hopefully represents the first example of a new paradigm for efficiently solving a variety of large-scale linear algebra problems. On a longer timeframe, the insights gained by studying VQMC realizations of NISQ algorithms on classical hardware will provide the basis for further quantum algorithm technology transfer as quantum hardware matures.
 
The paper is organized as follows. We compare and contrast in section \ref{vqa_vmc_section} the VQA-based and VQMC-based approaches for identifying the ground state of a quantum Hamiltonian. Section \ref{vqls_section} describes the Variational Quantum Linear Solver (VQLS) and how it encodes the solution of a linear system into the ground state of some Hamiltonian. Section \ref{vnls_section} introduces the Variational Neural Linear Solver (VNLS), a demonstrative example for producing fully-classical machine learning algorithms by applying VQMC techniques to VQA-inspired objectives. In section \ref{experiments_section}, we apply the VNLS to simple linear systems previously used in demonstration of the VQLS, before providing in section \ref{future_work_section} some general suggestions for how this area may be further developed. Many of the finer details in this paper are reserved for the appendix.


\section{VQAs and VQMC}
\label{vqa_vmc_section}

We now provide an overview of variational quantum algorithms for Hamiltonian ground state identification, presented within the context of potential classical simulation of these algorithms. We directly contrast VQAs with variational Monte Carlo techniques for accomplishing the same task.

\subsection{Variational Quantum Algorithms}

A variational quantum algorithm (VQA) is a hybrid quantum-classical algorithm simultaneously executed on a classical CPU and a noisy intermediate-scale quantum processing unit (QPU). VQAs are iterative: at each step, the CPU proposes a set of parameters to the QPU, which uses these values to execute a fixed sequence of parameterized quantum gates operating on a set of $n$ qubits. This sequence of gates, called an \textit{ansatz}, is kept shallow, meaning the number of gates in the sequence is polynomial with respect to $n$. The QPU passes information about the prepared state back to the CPU, which then performs some type of gradient-based update to the parameter set.

VQAs can solve a variety of tasks: the one most relevant to this paper is ground state identification for quantum Hamiltonians.
Consider the set $\{\ket{\psi_\theta}\in\mathbb{C}^{2^n} : \theta \in \mathbb{R}^d \}$ of all quantum state vectors of an $n$-qubit system that may be represented by different parameterizations of some shallow ansatz. We immediately observe, since the state vector $\ket{\psi_\theta}$ of our ansatz has dimension $2^n$, that the number of gates used to execute our ansatz is polylogarithmic with respect to the dimension of $\ket{\psi_\theta}$. Given a quantum Hamiltonian (Hermitian operator) $H$ on this $n$-qubit state space, the VQA seeks to find the parameterization $\theta$ minimizing the expected value of $H$ in state $\ket{\psi_\theta}$, which is derived from the Rayleigh-Ritz principle applied to a $2^n \times 2^n$ complex Hermitian matrix $H$,
\begin{equation}
    \label{rayleigh_quotient}
    L(\theta) = \frac{\langle \psi_\theta | H | \psi_\theta\rangle}{\langle \psi_\theta | \psi_\theta\rangle}
\end{equation}
The minimal expected value equals the smallest eigenvalue of $H$, which is achieved if and only if $\ket{\psi_\theta}$ is an associated eigenvector. Even under this specific type of VQA paradigm, one can hypothetically solve a variety of problems; all one has to do to make a problem solvable by a ground state identification VQA is to identify a Hamiltonian $H$ such that the solution of the desired problem is, in some way, encoded in its ground state. The Variational Quantum Linear Solver described in section \ref{vqls_section} provides an example of how one may construct such a Hamiltonian, but for the remainder of section \ref{vqa_vmc_section} we will only consider $H$ as an abstract Hamiltonian.

\subsection{Foundations of Classical Simulation of VQAs}
\label{vqa_simulability}

The overarching goal of this paper is to explore the utility of ``classicalizing'' variational quantum algorithms by replacing the QPU with a classical neural network, simulating all the quantum state preparation and expected value estimation tasks via Monte Carlo methods. Van den Nest \cite{nest2009simulating} introduces useful sufficient conditions under which quantum expectation values can be so estimated; in particular, the states and operators entering into quantum expectation values must satisfy conditions called classical tractability and efficiently computable sparsity, respectively.  A \textit{classically tractable} (CT) $n$-qubit state is one such that for any index $x\in [2^n]$ we may directly calculate the corresponding coordinate $\ip{x | \psi_\theta}$ with respect to the standard basis, and secondly, we may sample from the probability distribution on indices given by the Born rule, \begin{equation}
\label{born_rule}
\pi(x) = \frac{\abs{\ip{x|\psi_\theta}}^2}{\ip{\psi_\theta|\psi_\theta}}.\end{equation}
The normalization constant introduces a slight ambiguity of terminology, as quantum states are generally considered normalized, and true classical simulation of a CT state should maintain this property. However, it is certainly possible to model a CT state using an unnormalized vector, in which case one may approximate sampling from the corresponding Born distribution using Monte Carlo techniques. In this case, the Born rule includes this normalization constant, and it is this formulation of the rule that we follow in this paper.


We also require that the Hamiltonian $H$, which has dimension $2^n\times 2^n$, is \textit{efficiently computable}, meaning it obeys the following sparsity constraint: given any row $x\in [2^n]$ of $H$, we may calculate the column indices and values of all nonzero entries of row $x$ of $H$ in polynomial time with respect to $n$. Since $H$ is Hermitian, it is true that this property holds for all rows of $H$ if and only if it holds for all columns as well. It is clear that for $H$ to be efficiently computable, the number of nonzero entries in each row of $H$ must be polynomial in $n$; it is not, however, necessary that the number of \textit{all} nonzero entries in $H$ is polynomial in $n$. As we discuss in further detail in section \ref{vnls_algorithm_section}, the tensor product of local Pauli operators is efficiently computable, and the sum of a small---polynomial with respect to $n$---number of efficiently computable operators is efficiently computable.

\subsection{Variational Quantum Monte Carlo}
\label{vqmc_overview}
We now give a description of the VQMC algorithm, which provides the foundation for converting VQAs into fully classical algorithms. A brief description of how to implement this algorithm may be found in \ref{VQMC_implementation}. Consider a (possibly unnormalized) parameterized family of vectors $\ket{\psi_\theta}$ residing in a $2^n$-dimensional complex vector space. Let $\psi_\theta(x) := \ip{x| \psi_\theta}$ denote the components of $\psi_\theta$ relative to the standard orthornormal basis $\{ \ket{x} : x \in [2^n]\}$ of $\mathbb{C}^{2^n}$. Once normalized, $\ket{\psi_\theta}$ represents one potential quantum state in the state space of an $n$-qubit system. The starting point of the VQMC algorithm is the observation that the Rayleigh quotient for a Hermitian operator $H \in \mathbb{C}^{2^n \times 2^n}$ acting on this state space can be expressed in the following probabilistic form,
\begin{equation}
    \label{vmc_formulation}
    \frac{\bra{\psi_\theta}H\ket{\psi_\theta}}{\ip{\psi_\theta|\psi_\theta}} = \underset{x \sim \pi_\theta}{\mathbb{E}}\bracket{l_\theta(x)} \enspace \quad \quad l_\theta(x) :=\frac {\paren{H\psi_\theta}(x)} {\psi_\theta(x)} \enspace ,
\end{equation}
where the function $l_\theta(x)$ is referred to---for historical reasons---as the \textit{local energy}, and
where the distribution $\pi_\theta$ follows (\ref{born_rule}).
Equation (\ref{vmc_formulation}) follows from simple manipulations given in \ref{appendix_local_energy_derivation}.
The variance of the local $l_\theta(x)$ obeys the identity
\begin{equation}
\label{variance_eq}
    \var_{x\sim\pi_\theta}\big(l_\theta(x)\big) := \underset{x\sim\pi_\theta}{\mathbb{E}} \big[ \big|l_\theta(x) - L(\theta)\big|^2 \big] = \frac{\bra{\psi_\theta}H^2 \ket{\psi_\theta}}{\ip{\psi_\theta|\psi_\theta}} - \left[\frac{\bra{\psi_\theta} H \ket{\psi_\theta }}{\ip{\psi_\theta|\psi_\theta}}\right]^2 \enspace .
\end{equation}
A remarkable consequence of (\ref{variance_eq}) is that the variance of $l_\theta(x)$ approaches zero if $\psi_\theta$ approaches any eigenvector of $H$. This has the practical implication that convergence of the algorithm can be assessed by gathering statistics for the local energy.

The local energy $l_\theta(x)$ can be efficiently computed provided that the non-zero entries of the vector $H\psi_\theta$ can be extracted in a time which is polynomial in $n$, which is true if $\ket{\psi_\theta}$ is a CT state and $H$ is an efficiently computable Hamiltonian, as defined in section \ref{vqa_simulability}.

The VQMC procedure relies on the approximation of the Rayleigh quotient in (\ref{vmc_formulation}) by averaging batches sampled according to $\pi_\theta$ using Markov Chain Monte Carlo. More specifically, VQMC optimizes (\ref{Rayleigh}) by using a variant of stochastic gradient descent called stochastic reconfiguration \cite{sorella_aps98}, which is similar to stochastic natural gradient descent. In natural gradient descent, the parameter set $\theta$ is updated according to the rule $\theta\gets \theta - \gamma I^{+}(\theta)\nabla L(\theta)$, where $\gamma$ is the learning rate, $\nabla L(\theta)$ is the loss function gradient, and $I^{+}(\theta)$ represents the Moore--Penrose pseudoinverse of the Fisher information matrix at $\theta$, encoding the geometry of the optimization landscape near $\theta$. 

In the case that $\psi_\theta$ is a real vector---a slight simplification of the complex case---the gradient of (\ref{vmc_formulation}) and the Fisher information matrix at the parameterization $\theta$ of $\psi_\theta$ are given, respectively, by
\begin{equation}
    \nabla L(\theta) = \underset{x \sim \pi_\theta}{\mathbb{E}} \big[\big(l_\theta(x) - L(\theta) \big) \nabla_\theta \log |\psi_\theta(x)|\big], \; I(\theta) = \underset{x \sim \pi_\theta}{\mathbb{E}} \big[\nabla_\theta \log \pi_\theta(x) \otimes \nabla_\theta \log  \pi_\theta(x)\big].
\end{equation}
In practice, the gradient and Fisher information matrix are estimated stochastically---the gradient is estimated using batches of local energies sampled from $\pi_\theta$, and the Fisher information is estimated by computing sample covariances between the logarithmic partial derivatives $\frac{\partial}{\partial \theta_i} \paren{\log \abs{\psi_\theta(x)}}, \; x\sim \pi_\theta$ of $\psi_\theta$ with respect to $\pi_\theta$. 

\section{VQLS}
\label{vqls_section}
This section is dedicated to the Variational Quantum Linear Solver (VQLS) \cite{bravo2019variational}. We give a high-level description of the VQLS in section \ref{vqls_subsection}, followed by some common methods and examples for assessing its performance.

\subsection{The Variational Quantum Linear Solver}
\label{vqls_subsection}
The VQLS is a VQA designed to target linear systems in exponentially high dimensions of the form
\begin{equation}
    \label{vqls_problem_form}
    A\ket{x} \propto\ket{b}
\end{equation}
where $A$ denotes an invertible $2^n \times 2^n$ complex matrix and $\ket{b} \in \mathbb{C}^{2^n}$ is a nonzero vector. The output of the algorithm is an approximation of the quantum state equal to the normalization of the solution vector $A^{-1} \ket{b}$. For simplicity of exposition we assume that $A = A^\dag : = (\overline{A})^{\rm t}$ is a Hermitian matrix representing a quantum observable of the system. The original formulation of the VQLS merely presumes that $A$ is invertible, for which a more general formulation of the solver is given in \cite{bravo2019variational}. Furthermore it is demonstrated in \cite{harrow2009quantum} that one may always embed a non-Hermitian matrix on an $n$-qubit space within a Hermitian matrix using one additional ancilla qubit; this formulation is given in section \ref{appendix_matrix_embedding} of the appendix. Given a linear system of the form (\ref{vqls_problem_form}), consider the projection $P_b^{\perp}$ onto the subspace orthogonal to the vector $|b\rangle$,
\begin{equation}
  \quad \quad P_b^\perp := \mathbbm{1} - P_b \enspace , \quad \quad P_b := \frac{|b\rangle \langle b |}{\langle b | b \rangle}, \quad \quad H = A^\dag P_b^\perp A,
    \label{vqls_objective}
\end{equation}
with $H$ being the resulting Hamiltonian for which the VQLS seeks to minimize \eqref{rayleigh_quotient}.
This minimization occurs if and only if $\ket{\psi_\theta}$ is an eigenvector associated with the smallest eigenvalue of $H$, which in this case is 0. In particular, as $A$ is assumed to be invertible, the eigenspace associated with 0 is one-dimensional. Therefore for each $\theta$, equation (\ref{rayleigh_quotient}) gives a nonnegative value, equalling zero if and only if $\ket{\psi_\theta}\propto A^{-1}\ket{b}$; section \ref{error_meas} further shows exactly how the loss value bounds the accuracy of the VQLS.
\subsection{Error Bounding for the VQLS}
\label{error_meas}

Modulo a few scaling factors, the square root of the VQLS loss $L(\theta)$ gives an upper bound on the trace distance between the current model state $\ket{\psi_\theta}$ and the true solution state $\ket{A^{-1}b}$; more specifically, we find (after correcting a missing factor of $\Vert A \Vert$ in \cite{bravo2019variational}) that 
\begin{equation}
\operatorname{dist}_{\operatorname{Tr}}(\psi_\theta,A^{-1}b) = \frac 1 2 \operatorname{Tr}\paren{\sqrt{\big(\ketbra{\psi_\theta}{\psi_\theta} - \ketbra{A^{-1}b}{A^{-1}b}\big)^2}}\leq \frac{\kappa \sqrt{L(\theta)}}{\norm{A}},\end{equation} where $\kappa$ is the condition number of $A$ and $\norm{A}$ is its operator norm.
The trace distance formulation given here only holds for pure quantum states, which suffices for our purposes since our proposed VQLS-inspired solver (the VNLS) uses pure states. 
Another useful metric for comparing two quantum states is their \textit{fidelity}, which for two pure states equals the square of the cosine of the angle between their corresponding state vectors; in particular, identical quantum states have a fidelity of 1, and orthogonal quantum states have fidelity of 0. We discuss the trace distance bound, alongside how it implies an equivalent bound on the fidelities between these two quantum states, in appendix \ref{appendix_error_meas}.

\subsection{The Ising-Inspired VQLS Problem}
\label{comparison_analysis}

As a benchmark for assessing the performance of the VQLS, the authors of \cite{bravo2019variational} introduced an example problem which they term the Ising-inspired VQLS problem. In particular, given an $n$ qubit system and a condition number $\kappa$ they define  \begin{equation}A = \frac 1 \zeta \paren{\sum_{j=1}^n X_j + 0.1\sum_{j=1}^{n-1} Z_jZ_{j+1} + \eta I},\end{equation}
where $X_j$ and $Z_j$ denote the Pauli-$X$ and Pauli-$Z$ operators applied locally to qubit $j$, and $\zeta$ and $\eta$ are scaling factors specifically chosen such that $A$ has largest eigenvalue $1$ and smallest eigenvalue $\frac 1 \kappa$. The vector $\ket{b}$ is taken to be the equal superposition state on $n$ qubits given by $\ket{b} = |+\rangle^{\otimes n}$ where $|+\rangle =(1/\sqrt{2})(|0\rangle + |1\rangle)$. One noteworthy property of this problem is that the target $\ket{b}$ and solution $\ket{A^{-1}b}$ states become indistinguishable as the problem size increases. We give a further description and analysis of this phenomenon in appendix \ref{appendix_comparison_analysis} and \ref{appendex_ising_problem}.

\section{VNLS}
\label{vnls_section}
In this section we describe an instantiation of VQMC that we term the variational neural linear solver (VNLS). We focus on a linear system solver for concreteness, but the general strategy theoretically generalizes to a wide range of linear algebra problem, such as the SVD. The VNLS targets linear systems $A\ket{x}\propto\ket{b}$ of the same form as in section \ref{vqls_subsection}. Analogous to the VQLS, we consider the space $\{\ket{\psi_\theta}\in\mathbb{C}^{2^n} : \theta \in \mathbb{R}^d \}$ of parameterized vectors representing some collection of $n$-qubit  CT states. Given a $2^n \times 2^n$ Hermitian matrix $A$ and nonzero vector $\ket{b}$, we consider the VQLS objective
\begin{equation}
	\label{Rayleigh}
	L(\theta) :=
	\frac{\ip{\psi_\theta\abs{ AP_b^{\perp}A}\psi_\theta}}{\ip{\psi_\theta|\psi_\theta}} \enspace.
\end{equation} Just as is the case for VQLS, we have that $L(\theta)\geq 0$, with minimum value equal to $0$. As an alternative to preparing and storing $\ket{\psi_\theta}$ on an actual quantum device, we will take the approach of recasting the optimization problem into a form amenable to the variational quantum Monte Carlo, akin to the description given in section \ref{VQMC_implementation}.

\subsection{Applying VQMC to the linear systems problem}
\label{vnls_algorithm_section}
Here we give a formulation of the VNLS, alongside a description of how it may be efficiently implemented. The key observation is that VNLS is an instantiation of the VQMC using (\ref{vqls_objective}) as the input Hermitian Hamiltonian. As shown in \ref{appendix_vnls_objective_derivation}, we may apply the VQMC objective to this choice of $H$ and obtain a stochastic objective analogous to (\ref{vmc_formulation}), with a new local energy $l_\theta(x)$ defined by
\begin{equation}
    \label{vnls_local_energy}
	l_\theta(x) :=
	\frac 1 {\psi_\theta(x)}\paren{\big(A^2\psi_\theta\big)(x)-\big(Ab\big)(x) \, \underset{{x'\sim \beta}}{\mathbb{E}}\bracket{\frac{\big(A\psi_\theta\big)(x')}{b(x')}}} \enspace .
	\end{equation}
The expectation within (\ref{vnls_local_energy}) is taken with respect to the distribution $\beta(x)=\abs{b(x)}^2/\ip{b|b}$.
Thus we may approximate both the objective (\ref{Rayleigh}) and its gradient by estimating the expectation values over $\pi_\theta$ and $\beta$ using batches of data obtained using MCMC or other sampling methods. Algorithm \ref{vnls_alg} in \ref{appendix_vnls_objective_derivation} depicts this exact procedure and how it differs from \ref{VQMC_implementation}, traditional VQMC. For simplicity, we use the same batch size when sampling from $\pi_\theta$ and $\beta$, but doing so is not necessary in general.

The variational formulation given above places no additional theoretical restrictions on the structure of $A$ and $\ket{b}$, but some practical considerations have to be made in accordance with the more general Hamiltonian restrictions required for VQMC. The computational complexities of calculating $\paren{A^2 \psi_\theta}(x)$, $\paren{Ab}(x)$, and $\paren{A\psi_\theta}(x)$ grow with the number of nonzero entries in row $x$ of $A$, which can be as large as $2^n$. Similarly, the computational complexity of storing and sampling from the distribution $\beta$ grows with the number of nonzero entries of $\ket{b}$. For these reasons, we must require that $A$ is efficiently computable and that $\ket{b}$ is a CT state.

We ensure $A$ is efficiently computable in the following way. We observe that $A$, as a $2^n\times 2^n$ Hermitian matrix, may be expressed as a real linear combination of products of local Pauli operators $X$, $Y$, and $Z$. Our matrix $A$ will be efficiently computable so long as the number of terms in this sum is kept sufficently small. A description of what this decomposition entails is given in \ref{appendix_matrix_docmposition}. There are many different approaches for ensuring that $\ket{b}$ is CT, but for the purposes of this paper we simply store the nonzero elements of $\ket{b}$ directly and require that $\ket{b}$ either be sufficiently sparse or restrict our testing to problems of sufficiently small dimension.


An important detail to keep in mind when training the VNLS using the natural gradient is that scaling either the matrix $A$ or vector $b$ by any nonzero constant does not alter the problem in any way that the VNLS can meaningfully recognize, and does not provide any meaningful advantage in performance or accuracy.  Detailed explanations of these results are given in section \ref{appendix_vnls_scaling} of the appendix.

\section{Experimental Results}
\label{experiments_section}

As a benchmark test, we apply the VNLS to the Ising-inspired problem discussed in \ref{comparison_analysis}. We fix $\kappa=10$ and we take $\ket{b}$ to have all its entries equal to 1; while VQLS uses the normalized form of $\ket{b}$, scaling $\ket{b}$ does not influence the performance of VNLS in any significant way, as discussed in \ref{vnls_algorithm_section}.




\subsection{Real RBMs}
\label{real_rbm}
\begin{figure}
    \centering
    \includegraphics[width=\linewidth]{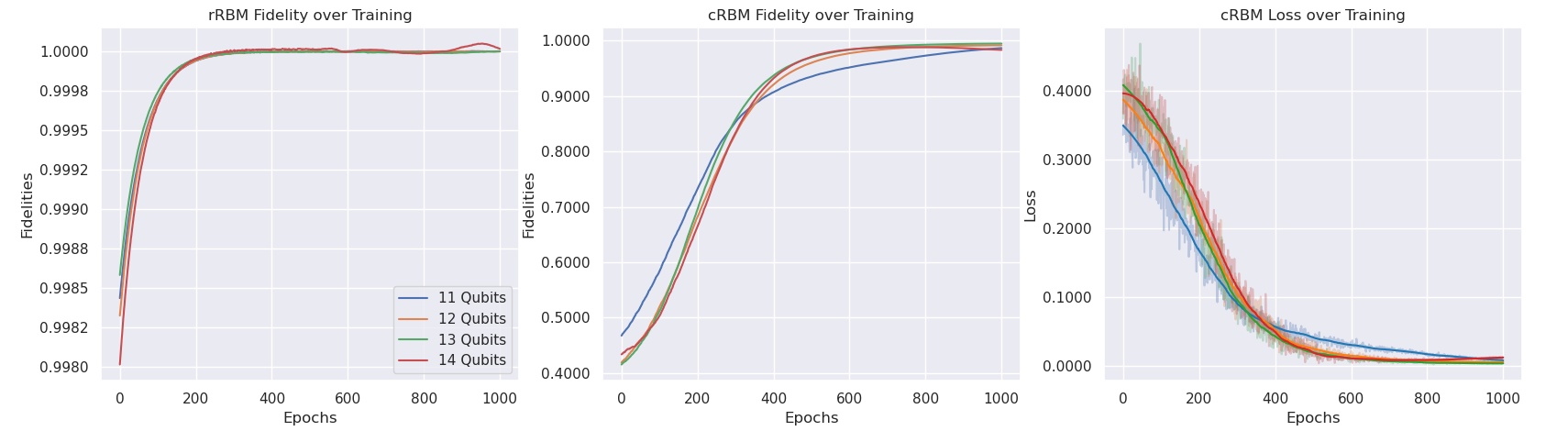}
    \caption{\small {\em Fidelities over training for a few $\kappa = 10$ Ising-inspired VQLS problems. Left: fidelities for training with a real RBM, which initializes closely to the true solution in all tested cases. Center: fidelities for training with a complex RBM. Right: evaluation loss curves for the same complex RBM tests shown in the center plot. For nontrivial $(n\geq 10)$ problem sizes, the complex RBM loss plots inversely correlate with the corresponding increases in fidelity, as expected.}}
    \label{fig:fidelity_loss_plots}
\end{figure}

Initializing the weights of a real RBM by sampling from a zero-mean Gaussian distribution with sufficiently small variance will, with high probability, place the corresponding state $\ket{\psi_\theta}$ in a small neighborhood of $\ket{b}$ with respect to the trace distance from \ref{error_meas}. As discussed in section \ref{comparison_analysis}, since $\ket{b}$ lies in a neighborhood of $A^{-1}\ket{b}$, we observe the Ising-inspired problem is trivially easy for a real RBM to learn, as shown in the left-most image in figure \ref{fig:fidelity_loss_plots}; this figure depicts, for qubit numbers 11 through 14, the fidelity between $\ket{\psi_\theta}$ and $A^{-1}\ket{b}$ over training. In all observed cases, the RBM initializes with very high fidelity, and only improves from there. These results are comparable with those obtained on larger problem sizes with analogous architecture.

These tests were done for $\kappa = 10$, where the RBM was trained with SR using a learning rate of $0.005$, over $1000$ epochs, with a batch size of $1024$ VQMC samples. Local energy samples were drawn from $8$ Monte Carlo chains. While directly calculating the fidelity between $\psi_\theta$ and $A^{-1}\ket{b}$ to assess performance is not viable at scale due to memory concerns, plotting this fidelity over the training period for small problems proves a better way for visually evaluating the model's performance than directly observing the stochastic loss plots. As the RBM initializes at a point with a small loss value, the noise in the stochastic loss makes it difficult to visually discern that the loss is decreasing over time from the loss graph alone. Though it does not harm the underlying point, we should note that the curve for the 14 qubit case is exhibiting roundoff errors, as it is impossible for the fidelity of two quantum states to be greater than 1.

\subsection{Complex RBMs}
In addition to the fact that a quirk of the Ising-inspired VQLS problem allows the real RBM to essentially initialize itself at the true solution state, real RBMs are likely not expressive enough to be of much use for the VNLS in practice, as they only learn quantum state vectors with entirely positive entries; there are certainly many linear systems of practical interest where the entries of the solution vector do not share the same sign. Furthermore, since quantum state vectors have complex entries in general, it is natural to explore the complex RBM (cRBM) for use in the VNLS.

Testing of the cRBM VNLS was performed using the $\kappa=10$ Ising-inspired problem with the same hyperparemeter configuration used in \ref{real_rbm}, yielding interesting qualitative results that would likely be more indicative of the VNLS's general behavior, in comparison with the somewhat trivial example of applying the real RBM (rRBM) to the Ising-inspired problem.

Unlike for rRBMs, the cRBM generally does not initialize in a state close to the solution. However, we may observe qualitatively that the variance between sampled losses of nearby epochs decreases over the course of training, as expected. The second observation is that the cRBM exhibits more heterogeneous learning behavior at small problem sizes ($n\leq 10$), occasionally struggling to learn the solution to the same standard of accuracy. We do not consider this a significant issue for two reasons: lowering the learning rate and training for more epochs appears to mitigate this issue somewhat, and moreover, just like the VQLS, the VNLS is not intended to be used for problem sizes suitable for standard numerical matrix inversion methods; we are primarily interested in its behavior at scale.



At larger problem sizes, the VNLS exhibits more uniform behavior, as demonstrated by the center and right-most images in figure \ref{fig:fidelity_loss_plots}. For the right-most image, depicting loss plots, we have smoothed the loss curves using a Savitzky--Golay filter, which better reveals the general trend. The translucent, noisy curves represent the true sampled loss plots; here, we can also observe the efficacy of the algorithm from the fact that the variances of nearby sample loss values decrease during training, which is a result of equation \ref{variance_eq}. Such behavior indicates that, relative to problem size, the optimization landscape of the Ising-inspired problem becomes easier for the VNLS to navigate. One likely explanation is that the deliberate fixing of the condition number across all problem sizes---which does not generally occur for matrices derived from physical applications---contributes to this phenomenon.

\begin{figure}
    \centering
    \includegraphics[width=\linewidth]{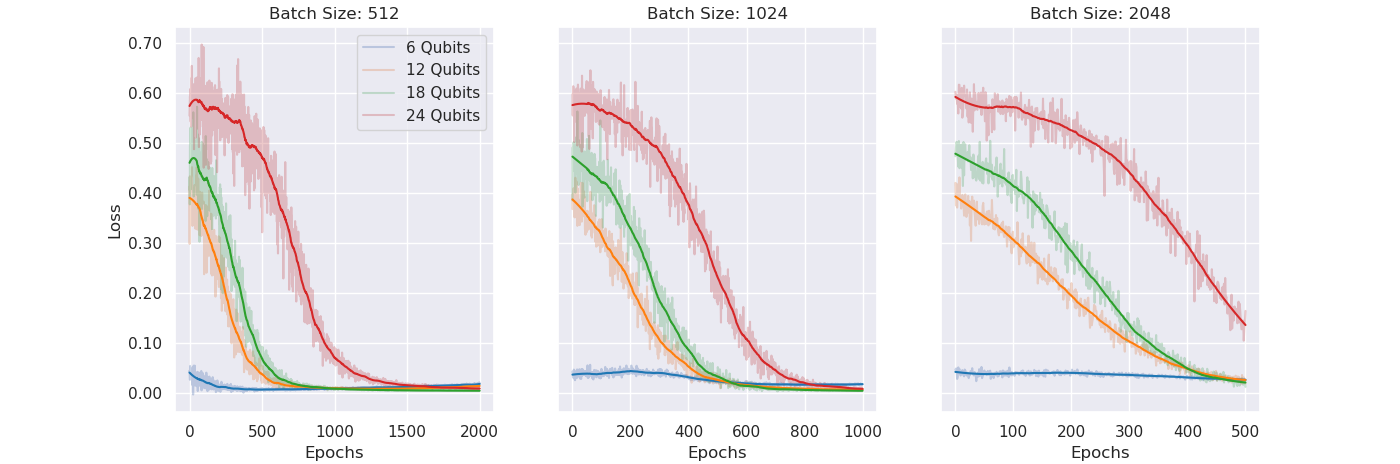}
    \includegraphics[width=\linewidth]{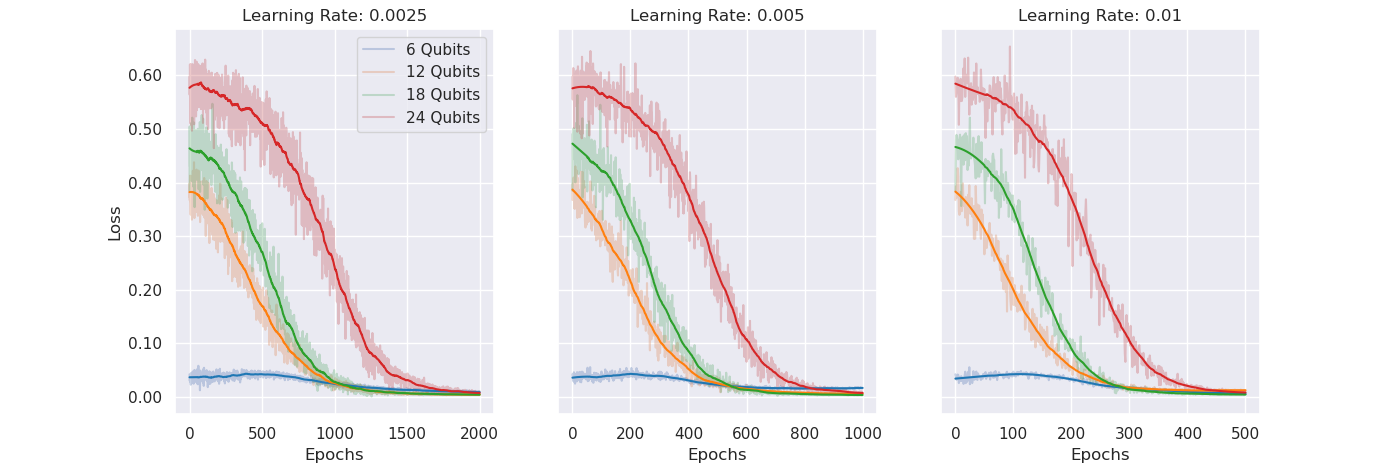}
    \caption{{\small\em Top row: effect of changing batch size on cRBM VNLS. Bottom row: effect of changing the learning rate. The epoch number is proportionately scaled so the graphs are qualitatively comparable.}}
    \label{batch_comparison}
\end{figure}

Figure \ref{batch_comparison} demonstrates the effect that altering the batch size or learning rate has on the performance of the VNLS. Batch size was varied between 512, 1024, and 2048 samples per epoch. In order to make the loss plots more comparable qualitatively, epoch numbers were scaled so that the total number of VQMC samples drawn remains fixed. We observe that while increasing the batch size does improve both accuracy and rate of convergence, this improvement does not scale linearly with the change in batch size. For instance, doubling the batch size from 512 to 1024 allow us to achieve comparable accuracy in half the time, though the VNLS requires a larger fraction of the training time to achieve it. in contrast, doubling the batch size from 1024 to 2048 does not halve the required training time in quite the same way.

For larger problem sizes, solely altering the learning rate within reasonable tolerances does not appear to affect performance of the cRBM VNLS in any qualitatively significant way beyond the expected commensurate change in the rate of convergence, as demonstrated by figure \ref{batch_comparison}. Here the number of training epochs was adjusted inversely in accordance with the change in learning rate. Across all larger ($n>10$) problem sizes tested, these learning rates were able to capture the same accuracy and general learning trend. It should be noted that for the smallest problem sizes, decreasing the learning rate did help to both improve accuracy and smooth out both the loss curve. These observations corroborate the idea that larger Ising-inspired problems are easier for the VNLS to solve.

\section{Future Work}
\label{future_work_section}

Through the application of the VNLS to solving linear systems, we have demonstrated a new paradigm for adapting VQAs to a wholly classical architecture by simulating the quantum circuit component with a neural network. There are a few natural points of expansion for this general area. Firstly, the choice of using an RBM for the neural network component is not vital to the algorithm itself, and there are no theoretical or empirical guarantees that RBMs are best suited for this specific task: one could instead leverage the recent success of autoregressive neural network quantum states \cite{sharir2020deep} or alternatively flow-based modeling for infinite-dimensional linear systems. Recent work in this direction indicates that it presents a scalable, practical alternative to RBMs \cite{zhao2021overcoming}. Another possible point of expansion would be to develop VQMC-based methods for performing other linear-algebraic operations in high dimensions, such as SVD or QR decomposition, by constructing Monte Carlo estimators based on recent developments in the VQA literature \cite{bravo2020quantum}. In addition to providing an ongoing benchmark for assessing quantum computational advantage of VQAs, continued research on VQA-inspired classical algorithms might provide valuable insight into the structure of quantum states for which VQAs might yield quantum advantage.



\begin{ack}
Authors acknowledge support from NSF under grant DMS-2038030.
\end{ack}

\bibliographystyle{plainnat}
\bibliography{references}


\appendix

\section{Appendix}
\subsection{Derivation of the Local Energy}
\label{appendix_local_energy_derivation}
Expanding $\psi_\theta$ in the standard orthonormal basis, using anti-linearity in the first argument gives
\begin{align}
    \frac{\bra{\psi_\theta}H\ket{\psi_\theta}}{\ip{\psi_\theta|\psi_\theta}}
    & = \sum_{x \in [2^n]} \frac{\overline{\psi_\theta(x)}}{\ip{
    \psi_\theta|\psi_\theta}} (H\psi_\theta)(x) = \sum'_{x \in [2^n]} \frac{\overline{\psi_\theta(x)}}{\ip{
    \psi_\theta|\psi_\theta}} (H\psi_\theta)(x)
\end{align}
where prime in the second summation indicates restriction to the nonzero entries of $\psi_\theta$. Dividing and multiplying the summand by $\psi_\theta(x)$, one obtains the desired result:
\begin{equation}
    \frac{\bra{\psi_\theta}H\ket{\psi_\theta}}{\ip{\psi_\theta|\psi_\theta}}
     =\sum'_{x \in [2^n]} \frac{|\psi_\theta(x)|^2}{\ip{\psi_\theta|\psi_\theta}} \frac{(H\psi_\theta)(x)}{\psi_\theta(x)} = \underset{x \sim \pi_\theta}{\mathbb{E}}\bracket{l_\theta(x)} \enspace .
\end{equation}
\subsection{Embeddings of Non-Hermitian Matrices}
\label{appendix_matrix_embedding}

As demonstrated in \cite{harrow2009quantum}, when presented with a non-Hermitian $2^n\times 2^n$ matrix $A$, corresponding with an $n$-qubit state space, one can always embed $A$ into a Hermitian matrix at the cost of one additional qubit, yielding the following modified linear system,
\begin{equation}
\begin{bmatrix}
0 & A\\
A^\dagger & 0\\ 
\end{bmatrix}\ket{x} = \begin{bmatrix} \ket{b}\\ 0\\\end{bmatrix}, \text{\quad \quad whose solution is given by \quad\quad} \begin{bmatrix} 0 \\ A^{-1}\ket{b}\\ \end{bmatrix}.
\end{equation}

\subsection{Implementing VQMC}
\label{VQMC_implementation}
\begin{algorithm}
    \begin{algorithmic}
        \State \textbf{Input:} Hamiltonian $H$, quantum state neural network $\psi_\theta$, batch size $k$, learning rate $\gamma$
        \State Initialize $\theta$
        \While {not done}
            \State Obtain $k$ samples $x_1,\dots,x_k$ from $\pi_\theta$
            \For {$i=1,\dots,k$}
                \State Obtain $(H\psi_\theta)(x)$ by multiplying the nonzero entries of row $x$ of $H$ with the corresponding
                \State entries of $\psi_\theta$, and summing the results
                \State Compute local energy $l_\theta(x_i)$
                \State Compute logarithmic gradients $\nabla_\theta \log\abs{\psi_\theta(x)}$ using automatic differentiation
            \EndFor
            \State Estimate objective using sample mean $\hat{L}(\theta) = \frac 1 k \sum_{i=1}^k l_\theta(x_i)$
            \State Estimate gradient and Fisher information using sample mean $\nabla\hat{L}(\theta)$ and covariance matrix $\hat{I}(\theta)$
            \State Perform parameter update $\theta \gets \textsc{Optimizer}\paren{\gamma, \theta, \nabla \hat{L}, \hat{I}}$
        \EndWhile
    
    \end{algorithmic}
    \caption{Variational Quantum Monte Carlo using Neural Networks}
    \label{VQMC_alg}
\end{algorithm}

Algorithm \ref{VQMC_alg} gives a succinct description of how VQMC may be implemented in practice. One may make a direct comparison between the high-level structures of VQMC and VQAs for Hamiltonian ground state identification. The essential difference between the two that in VQMC we assume $\ket{\psi_\theta}$ is a neural-network quantum state in the vein of what is described in section \ref{vqa_simulability}. For example, given an input $x\in [2^n]$ one can model $\psi_\theta(x)$ by the output of a restricted Boltzmann machine (RBM) with $n$ visible nodes; inputting the binary encoding of $x$ into the RBM returns component $\psi_\theta(x)$.

It should be noted here that VQMC---and, by extension, VNLS---is network-agnostic. Nothing in what follows will depend upon the use of RBMs, and there is good reason to consider more sophisticated architectures including autoregressive models, which are popular in natural language processing.

\subsection{VQLS Error Bound Discussion}
\label{appendix_error_meas}

Here we describe how the exact value of the objective function (\ref{vqls_objective}) informs us of the accuracy of the VQLS, even when the loss value is nonzero. The key idea behind this correspondence is given in \cite{bravo2019variational}, though there is a slight modification that must be made to the result, which we note.

The primary method for assessing and bounding the error of the VQLS is by using the trace distance between two quantum states, which is given for two pure states $\ket{\phi}$ and $\ket{\psi}$ by \begin{equation}\operatorname{dist}_{\operatorname{Tr}}(\phi,\psi) = \frac 1 2 \operatorname{Tr}\paren{\sqrt{\paren{\ketbra{\phi}{\phi} - \ketbra{\psi}{\psi}}^2}}.\end{equation} A more general definition of trace distance is unnecessary for our purposes since our own VQLS-inspired classical solver, the VNLS, does not model mixed quantum states, only pure ones. Under the assumption that $\norm{A}\leq 1$, it is stated in \cite{bravo2019variational} that \begin{equation}
\label{false_bound}
    \operatorname{dist}_{\operatorname{Tr}}(\psi_\theta,A^{-1}b)\leq \kappa \sqrt{L(\theta)},
\end{equation} where $\kappa$ is the condition number of $A$. As discussed in section \ref{vnls_algorithm_section}, scaling $A$ by a positive constant $c<1$ can make the right-hand side of (\ref{false_bound}) arbitrary small without actually changing the trace distance between the learned and true states, meaning this bound does not hold in its current form.

The discrepancy in the proof of this bound comes from an intermediate result taken from \cite{subasi2019linear}, which states that the smallest eigenvalue of the matrix $A^2$, which equals $\frac{\norm{A}^2}{\kappa^2}$, is bounded below by \begin{equation}\frac{\norm{A}^2}{\kappa^2}\geq \frac 1 {\kappa^2}\end{equation} when $\norm{A}\leq 1$. In this case, $\frac 1 {\kappa^2}$ is actually an upper bound, not a lower one. Adapting around this small error, we may still follow the rest of the argument given in \cite{bravo2019variational} to obtain a new error bound, \begin{equation}
    \operatorname{dist}_{\operatorname{Tr}}(\psi_\theta,A^{-1}b)\leq \frac{\kappa \sqrt{L(\theta)}}{\norm{A}},
    \label{true_bound}
\end{equation} which agrees with the scaling argument given above. This new bound does not require any restriction on $\norm{A}$, and is equivalent to saying that the trace distance between $\ket{\psi_\theta}$ and $A^{-1}\ket{b}$ is bounded above by the product of $\sqrt{L(\theta)}$ with the smallest singular value of $A$.

Alternately, since the VQLS ansatz $\ket{\psi_\theta}$ and the true solution $A^{-1}\ket{b}$ correspond with a pair of pure quantum states when normalized, an equivalent way of measuring the solver's error would be through using the fidelity between these two states, given here for two general states $\ket{\psi}$ and $\ket{\phi}$: \begin{equation}\operatorname{Fid}(\phi,\psi) = \frac{\abs{\ip{\phi|\psi}}^2}{\ip{\phi|\phi}\ip{\psi|\psi}}.\end{equation} The fidelity equals 1 if and only if one vector is a scalar multiple of the other, and 0 if the two vectors are orthogonal.

The two measures are essentially equivalent for pure states; more specifically, we have that \begin{equation}\operatorname{dist}_{\operatorname{Tr}}(\psi_\theta,A^{-1}b) = \sqrt{1 - \operatorname{Fid}(\psi_\theta,A^{-1}b)},\end{equation} from which it follows that any upper bound on the trace distance corresponds uniquely with a lower bound on the fidelity; thus the bound given in (\ref{true_bound}) could be used in either case.

\subsection{Ising-Inspired VQLS Problem Discussion}
\begin{figure}
    \centering
    \includegraphics[scale=0.5]{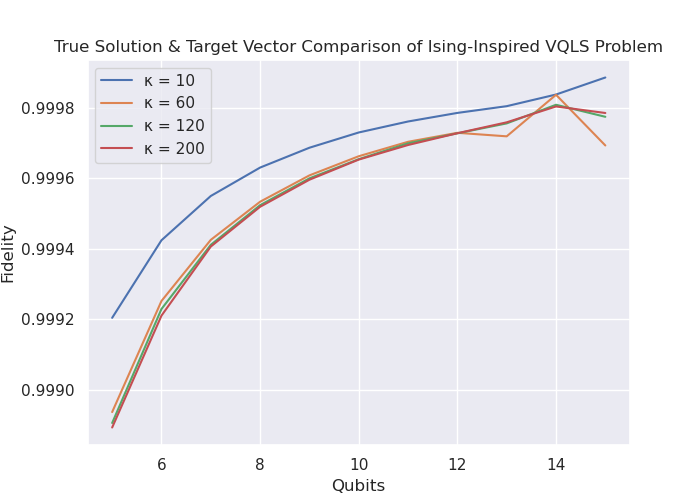}
    \caption{\small {\em Fidelity, with respect to problem size between target $\ket{b}$ and solution $A^{-1}\ket{b}$ vectors for the Ising-inspired VQLS problem.}}
    \label{fig:true_target_comparison}
\end{figure}
\label{appendix_comparison_analysis}

Here we give a more detailed description and analysis of the Ising-inspired VQLS problem from \cite{bravo2019variational}. Given a qubit number $n$ and a condition number $\kappa$, we consider the matrix \begin{equation}A = \frac 1 \zeta \paren{\sum_{j=1}^n X_j + 0.1\sum_{j=1}^{n-1} Z_jZ_{j+1} + \eta I},\end{equation} where $X_j$ and $Z_j$ refer to the Pauli-$X$ and Pauli-$Z$ operators applied locally to qubit $j$, and $\zeta$ and $\eta$ are chosen so that $A$ has largest eigenvalue $1$ and smallest eigenvalue $\frac 1 \kappa$. The vector $\ket{b}$ is taken to be the equal superposition state on $n$ qubits given by \begin{equation}\ket{b} = \paren{\frac 1{\sqrt 2}\begin{bmatrix} 1 & 1\\ \end{bmatrix}^T}^{\otimes n}\end{equation}

We observe that while $A$ is surely not the identity matrix, it does behave similarly to the identity when applied to $\ket{b}$ for nontrivial problem sizes. We first note that for a given problem size $n$ and condition number $\kappa$, we may take coefficients \begin{equation}\eta = n\paren{\frac{\kappa + 1}{\kappa - 1}}\text{ and }\zeta = n + \eta = \frac{2n\kappa}{\kappa - 1}.\end{equation} This choice is based on a heuristic observation that the largest and smallest eigenvalues of \begin{equation}\sum_{j=1}^n X_j + 0.1 \sum_{j=1}^{n-1} Z_j Z_{j+1}\end{equation} are approximately $n$ and $-n$, in cases for which the qubit number is small enough that these eigenvalues can be easily computed numerically.

The vector $\begin{bmatrix} 1 & 1\\ \end{bmatrix}^T$ is an eigenvector of the matrix $X$ with eigenvalue $1$, meaning that applying $X$ locally to any one of the $n$ qubits does not change the system state $\ket{b}$. Thus $\ket{b}$ is an eigenvector of $\sum_{j=1}^n X_j$ with eigenvalue $n$. It follows that the matrix $A$ perturbs $\ket{b}$ in the following way: 

\begin{equation}
    \label{VQLS_problem_perturbation}
    A\ket{b} = \ket{b} + \frac{0.05(\kappa-1)}{n\kappa} \sum_{j=1}^{n-1} Z_jZ_{j+1}\ket{b}.
\end{equation}

It follows that $A$, when applied to $\ket{b}$, will perturb each entry of $\ket{b}$ by at most $2^{-\frac n 2}\paren{0.05}$, an explanation for which is given in section \ref{appendex_ising_problem} of the appendix.

Just as $A$ perturbs the vector $\ket{b}$, it appears that $A^{-1}$ acts on $\ket{b}$ in a similar manner. Experimentally, we have found that for problem sizes of at least five qubits, the fidelity between $\ket{b}$ and $A^{-1}\ket{b}$ is significantly close to 1. Figure \ref{fig:true_target_comparison} demonstrates that this fidelity grows larger with problem size, and that this trend persists even as the condition number $\kappa$ increases.

Seeking to explain this phenomenon, we examine the Euclidean distance between $\ket{b}$ and $A^{-1}\ket{b}$. We observe that

\begin{equation}
    \norm{\ket{b} - A^{-1}\ket{b}}^2 \leq \frac{0.0025(\kappa - 1)^2}{n}\paren{\frac{n-1}{n}}\rightarrow 0 \text{ as } n\rightarrow\infty,
\end{equation}
for which a proof is also given in \ref{appendex_ising_problem}. While it is possible that this upper bound is not tight, it is nonetheless clear that for a fixed choice of $\kappa$, the vectors $\ket{b}$ and $A^{-1}\ket{b}$ become indistinguishable by the VQLS and the VNLS as the qubit number becomes arbitrarily large. Figure \ref{fig:true_target_comparison} demonstrates this phenomenon in practice.

\subsection{Correspondence between Target and Solution States for Ising-Inspired VQLS Problem}
\label{appendex_ising_problem}

We recall equation \ref{VQLS_problem_perturbation}, which states for the Ising-inspired VQLS problem that

\begin{equation}A\ket{b} = \ket{b} + \frac{0.05(\kappa-1)}{n\kappa} \sum_{j=1}^{n-1} Z_jZ_{j+1}\ket{b}.\end{equation}

We now show how exactly the local $Z$ term in \ref{VQLS_problem_perturbation} perturbs the entries of $\ket{b}$. The $Z$-matrix maps $\begin{bmatrix} 1 & 1\\ \end{bmatrix}^T$ to $\begin{bmatrix} 1 & -1\\ \end{bmatrix}^T$, so for each $i= 0,\dots,n-1$, applying $Z$ locally to qubits $i$ and $i+1$ converts the state $\ket{b}$ into \begin{equation} 2^{-\frac n 2}\begin{bmatrix} 1\\ 1\\ \end{bmatrix}_1\otimes\dots\otimes \begin{bmatrix} 1\\ -1\\ \end{bmatrix}_i \otimes \begin{bmatrix} 1\\ -1\\ \end{bmatrix}_{i+1} \otimes \dots \otimes \begin{bmatrix} 1\\ 1\\ \end{bmatrix}_n\end{equation}

By exploiting the structure of the Kronecker product in exactly the same manner as is done in \ref{appendix_matrix_docmposition}, we may calculate specific coordinates of $Z_jZ_{j+1}\ket{b}$ as follows: consider entry $x\in\braces{0,1,\dots,2^{n-1}}$ of $Z_jZ_{j+1}\ket{b}$, represented in binary by $(x_1,\dots,x_n)$. The value of entry $x$ equals $2^{-\frac n 2}$ multiplied by the product, taken over all $k=1,\dots,n$ qubits, of entry $x_k$ of either $\begin{bmatrix} 1 & 1\\ \end{bmatrix}^T$ or $\begin{bmatrix} 1 & -1\\ \end{bmatrix}^T$, depending on whether or not $k \in\braces{j, j+1}$.

Thus we may observe that entry $x$ of $Z_jZ_{j+1}\ket{b}$ equals $2^{-\frac n 2}$ if $x_j$ and $x_{j+1}$ agree, and $-2^{-\frac n 2}$ if they do not. Since $\sum_{j=1}^{n-1}Z_{j}Z_{j+1}$ iterates over every pair of consecutive qubits, it follows that entry $x$ of $\sum_{j=1}^{n-1}Z_{j}Z_{j+1}\ket{b}$ equals the difference, scaled by $2^{-\frac n 2}$, between the number of consecutive binary digits in $(x_1,\dots,x_n)$ with the same value, and the number of consecutive binary digits with opposing values. It follows from (\ref{VQLS_problem_perturbation}) that applying $A$ to $\ket{b}$ perturbs each entry of $\ket{b}$ by at most $2^{-\frac n 2}\paren{0.05}$.

We now show the effect of $A^{-1}$ on $\ket{b}$. We observe that the square of the Euclidean distance between $A^{-1}\ket{b}$ and $\ket{b}$ may be bounded above as follows:

\begin{align}
    \nonumber
    \norm{\ket{b} - A^{-1}\ket{b}}^2 & = \norm{A^{-1}\paren{A\ket{b} - \ket{b}}}^2\\
    \nonumber
    & \leq \norm{A^{-1}}^2\norm{\paren{A\ket{b} - \ket{b}}}^2\\
    \nonumber
    & = \kappa^2 \norm{\frac{0.05(\kappa-1)}{n\kappa} \sum_{j=1}^{n-1} Z_jZ_{j+1}\ket{b}}^2\\
    \nonumber
    & = \frac{0.0025(\kappa - 1)^2}{n^2}\norm{ \sum_{j=1}^{n-1} Z_jZ_{j+1}\ket{b}}^2\\
    \nonumber
    & = \frac{0.0025(\kappa - 1)^2}{n^2}\ip{\sum_{j=1}^{n-1} Z_jZ_{j+1}b\bigg|\sum_{k=1}^{n-1} Z_kZ_{k+1}b}\\
    \nonumber
    & = \frac{0.0025(\kappa - 1)^2}{n^2}\sum_{j=1}^{n-1}\sum_{k=1}^{n-1}\ip{ Z_jZ_{j+1}b| Z_kZ_{k+1}b}\\
    & = \frac{0.0025(\kappa - 1)^2}{n^2}\sum_{j=1}^{n-1}\sum_{k=1}^{n-1}\ip{b\abs{ Z_{j+1} Z_jZ_kZ_{k+1}}b} \label{VQLS_bound_first_half}
\end{align}

Equation \ref{VQLS_bound_first_half} follows from the line preceding it by the fact that the local $Z$-operator is Hermitian. To complete the bound, we fix any $j,k\in \braces{1,\dots,n-1}$ and consider the inner product $\ip{b| Z_{j+1} Z_jZ_kZ_{k+1}b}$. Since $\ket{b}$ is the tensor product of $n$ copies of the vector $2^{-\frac 1 2}\begin{bmatrix}
1 & 1\\
\end{bmatrix}^T$, and the operator $Z_{j+1} Z_jZ_kZ_{k+1}$ acts locally on $\ket{b}$, it follows that we may decompose the inner product into \begin{equation}\ip{b\abs{ Z_{j+1} Z_jZ_kZ_{k+1}}b} = 2^{-n} \prod_{\ell=1}^n \begin{bmatrix}
1 & 1\\
\end{bmatrix}O_\ell \begin{bmatrix}
1\\
1\\
\end{bmatrix}, \label{inner_product_decomp}\end{equation} where for each $\ell$ the local operator $O_\ell$ is given by \begin{equation}O_\ell = \begin{cases}
I & \text{if } \ell \text{ does not equal any of } j,j+1,k,k+1\\
Z & \text{if } \ell \text{ equals exactly one of }j,j+1,k,k+1\\
Z^2 = I & \text{if } \ell \text{ equals both one of }j,j+1 \text{ and one of }k,k+1.\\
\end{cases}\end{equation}

Since the products \begin{equation}\begin{bmatrix}
1 & 1\\
\end{bmatrix}I \begin{bmatrix}
1\\
1\\
\end{bmatrix} = 1 + 1 = 2\text{ and }\begin{bmatrix}
1 & 1\\
\end{bmatrix}Z \begin{bmatrix}
1\\
1\\
\end{bmatrix} = 1-1=0,\end{equation} it follows from (\ref{inner_product_decomp}) that $\ip{b| Z_{j+1} Z_jZ_kZ_{k+1}b}$ equals $1$ if $j=k$ and $0$ otherwise. Therefore we have from (\ref{VQLS_bound_first_half}) that 
\begin{align}
    \nonumber
    \norm{\ket{b} - A^{-1}\ket{b}}^2 & \leq \frac{0.0025(\kappa - 1)^2}{n^2}\sum_{j=1}^{n-1}\sum_{k=1}^{n-1}\ip{b\abs{ Z_{j+1} Z_jZ_kZ_{k+1}}b}\\
    \nonumber
    & = \frac{0.0025(\kappa - 1)^2}{n^2}\sum_{j=1}^{n-1}1\\
    & = \frac{0.0025(\kappa - 1)^2}{n}\paren{\frac{n-1}{n}}\rightarrow 0 \text{ as } n\rightarrow\infty
\end{align}

\subsection{VNLS Objective Derivation}
\label{appendix_vnls_objective_derivation}
Evaluating the VQMC objective for this specific choice of $H$ one finds
\begin{align}
	L(\theta) & = \frac{\ip{\psi_\theta\abs{A^2-\frac{A\ketbra{b}{b}A}{\ip{b|b}}}\psi_\theta}}{\ip{\psi_\theta|\psi_\theta}} \enspace , \\
	& =\frac{\bra{\psi_\theta}A^2\ket{\psi_\theta}}{\ip{\psi_\theta|\psi_\theta}} - \frac{\bra{\psi_\theta} \bra{b} A\ket{\psi_\theta}A\ket{b}}{\ip{\psi_\theta|\psi_\theta}\ip{b|b}} \enspace , \\ & =\frac{\bra{\psi_\theta}A^2\ket{\psi_\theta}}{\ip{\psi_\theta|\psi_\theta}} - \paren{\frac{\bra{\psi_\theta} A\ket{b}}{\ip{\psi_\theta|\psi_\theta}}}\paren{\frac{\bra{b}A\ket{\psi_\theta}}{\ip{b|b}}} \enspace .
\end{align}
By similar manipulations as in section \ref{vqmc_overview}, we find that the VNLS objective function takes the probabilistic form
\begin{equation}
    L(\theta) = \underset{x \sim \pi_\theta}{\mathbb{E}}\bracket{l_\theta(x)} \enspace ,
\end{equation}
where the local energy is
\begin{equation}
	l_\theta(x) :=
	\frac 1 {\psi_\theta(x)}\paren{\big(A^2\psi_\theta\big)(x)-\big(Ab\big)(x) \, \underset{{x'\sim \beta}}{\mathbb{E}}\bracket{\frac{\big(A\psi_\theta\big)(x')}{b(x')}}} \enspace ,
	\end{equation}
where the distribution $\beta$ is defined by $\beta(x)=\abs{b(x)}^2/\ip{b|b}$. Algorithm \ref{vnls_alg} demonstrates the VNLS in full, highlighting its similarities and differences with traditional VQMC.

\begin{algorithm}
    \begin{algorithmic}
        \State \textbf{Input:} Matrix $A$, vector $\ket{b}$, quantum state neural network $\psi_\theta$, batch size $k$, learning rate $\gamma$
        \State Initialize $\theta$
        \While {not done}
            \State Obtain $k$ samples $x_1,\dots,x_k$ from $\pi_\theta$ and $x'_1,\dots,x'_k$ from $\beta$
             \For {$i=1,\dots,k$}
                \State Calculate $\frac{(A\psi_\theta)(x'_i)}{b(x'_i)}$
            \EndFor
            \State Estimate expected value w.r.t. $\beta$ using $\frac 1 k \sum_{i=1}^k \frac{(A\psi_\theta)(x'_i)}{b(x'_i)}$
            \For {$i=1,\dots,k$}
                \State Obtain $(A^2\psi_\theta)(x)$ by multiplying nonzero entries of row $x$ of $A^2$ with the corresponding
                \State entries of $\ket{\psi_\theta}$, summing the results
                \State Do the same for $(Ab)(x)$
                \State Construct local energy $\hat{l}_\theta(x_i)$ estimate
                \State Compute logarithmic gradients $\nabla_\theta \log\abs{\psi_\theta(x)}$ using automatic differentiation
            \EndFor
            \State Estimate objective using sample mean $\hat{L}(\theta) = \frac 1 k \sum_{i=1}^k \hat{l}_\theta(x_i)$
            \State Estimate gradient and Fisher information using sample mean $\nabla\hat{L}(\theta)$ and covariance $\hat{I}(\theta)$
            \State Perform parameter update $\theta \gets \textsc{Optimizer}\paren{\gamma, \theta, \nabla \hat{L}, \hat{I}}$
        \EndWhile
    
    \end{algorithmic}
    \caption{Variational Neural Linear Solver}
    \label{vnls_alg}
\end{algorithm}

\subsection{Equivalency of VNLS Problems Under Scaling}
\label{appendix_vnls_scaling}

We demonstrate here how the scaling of a VNLS problem, as described in section \ref{vnls_algorithm_section}, has no meaningful effect on the ability of VNLS to solve it using SR.

Recalling that the goal of VNLS is to find, for a given Hermitian matrix $A$ and vector $\ket{b}$, an unnormalized vector $\ket{x}$ such that $A\ket{x}$ is proportional to $\ket{b}$, we observe that if \begin{equation}A\ket{x}\propto \ket{b},\text{ then }c_1A\ket{x}\propto c_2 \ket{b}\end{equation} for any pair of nonzero scalars $c_1$ and $c_2$. For this reason, the solution set of any VNLS problem is invariant under scaling of both $A$ and $\ket{b}$. In practice, one may consider that scaling either $A$ or $\ket{b}$ by some positive value might improve the learning process in some cases, but we find that careful scaling of $A$ or $\ket{b}$ yields no unique advantage for VNLS performance under stochastic reconfiguration.

In particular, we find that keeping $A$, $\theta$, and $x$ fixed, scaling $\ket{b}$ by any positive constant $c$ does not change the VNLS objective $L(\theta)$ or its gradient. Similarly, keeping $\ket{b}$, $\theta$, and $x$ fixed, scaling $A$ by any positive constant $c$ changes the objective $L(\theta)$ and its gradient $\nabla L(\theta)$ to $c^2L(\theta)$ and $c^2\nabla L(\theta)$, which we now explain.

For a given VNLS problem $\braces{A,\ket{b}}$ and parameterized, unnormalized quantum state $\ket{\psi_\theta}$, we recall from (\ref{vnls_local_energy}) that the local energy of the VNLS Hamiltonian at basis state $x$ is given by \begin{equation}l_\theta(x) =
	\frac 1 {\psi_\theta(x)}\paren{\big(A^2\psi_\theta\big)(x)-\big(Ab\big)(x) \, \underset{{x'\sim \beta}}{\mathbb{E}}\bracket{\frac{\big(A\psi_\theta\big)(x')}{b(x')}}},\end{equation} with the VNLS objective being the expected value of $l_\theta(x)$ with respect to the distribution $\pi_\theta$ corresponding with state $\ket{\psi_\theta}$. Keeping $A$, $\theta$, and $x$ fixed, we observe that replacing $\ket{b}$ with $c\ket{b}$ for any positive scalar $c$ preserves the local energy: \begin{equation}\frac 1 {\psi_\theta(x)}\paren{\big(A^2\psi_\theta\big)(x)-\big(A(cb)\big)(x) \, \underset{{x'\sim \beta}}{\mathbb{E}}\bracket{\frac{\big(A\psi_\theta\big)(x')}{cb(x')}}}\end{equation} \begin{equation}=
	\frac 1 {\psi_\theta(x)}\paren{\big(A^2\psi_\theta\big)(x)-\frac c c\big(Ab\big)(x) \, \underset{{x'\sim \beta}}{\mathbb{E}}\bracket{\frac{\big(A\psi_\theta\big)(x')}{b(x')}}} = l_\theta(x).\end{equation} This replacement also has no effect on the distribution $\beta$, and consequently scaling the vector $b$ does not affect VNLS training in any way.
	
On the other hand, scaling $A$ while holding $\ket{b}$, $\theta$, and $x$ fixed does alter the local energy: \begin{equation}\frac 1 {\psi_\theta(x)}\paren{\big((cA)^2\psi_\theta\big)(x)-\big(cAb\big)(x) \, \underset{{x'\sim \beta}}{\mathbb{E}}\bracket{\frac{\big(cA\psi_\theta\big)(x')}{b(x')}}}\end{equation} \begin{equation}=
	\frac {c^2} {\psi_\theta(x)}\paren{\big(A^2\psi_\theta\big)(x)-\big(Ab\big)(x) \, \underset{{x'\sim \beta}}{\mathbb{E}}\bracket{\frac{\big(A\psi_\theta\big)(x')}{b(x')}}}= c^2l_\theta(x).\end{equation} It follows that our objective $L(\theta)=\underset{{x'\sim \pi_\theta}}{\mathbb{E}}\bracket{l_\theta(x)}$ then becomes $c^2L(\theta)$, hence the gradient of our objective is replaced with $c^2\nabla L(\theta)$.

 Under stochastic reconfiguration, solving the VNLS problem $\braces{cA, \ket{b}}$ using learning rate $\gamma$ is then tantamount to solving the original problem $\braces{A,\ket{b}}$ using learning rate $c^2\gamma$, with the caveat that the objective function is scaled by $c^2$ as well. This latter change notwithstanding, scaling $A$ does not affect SR in any way that could not similarly be achieved by changing the learning rate.

\subsection{Pauli Basis Decomposition of Hermitian Matrices}
\label{appendix_matrix_docmposition}
From this point onward we will treat the rows and columns of every matrix discussed as being indexed starting from 0. As discussed in section \ref{vnls_algorithm_section}, we may decompose any $2^n\times2^n$ Hermitian matrix $A$ into a linear combination

 \begin{equation}A = \sum_{i}c_i\paren{A_{i_1}\otimes\cdots\otimes A_{i_n}},\end{equation} where each $A_{i_{j}}$ is either the identity operator or one of the Pauli operators \begin{equation}X = \begin{bmatrix}
0 & 1\\
1 & 0 \\ \end{bmatrix}, Y = \begin{bmatrix}
0 & -i\\
i & 0\\ \end{bmatrix}, \text{ and } Z = \begin{bmatrix}
1 & 0\\
0 & -1\\ \end{bmatrix}.\end{equation} Here the notation $\otimes$ refers to the Kronecker product \begin{equation} \label{kronecker}
A \otimes B = \begin{bmatrix}
A_{00}B & A_{01}B\\
A_{10}B & A_{11}B \end{bmatrix}, \end{equation}

which may be applied recursively to produce each summand in $A$. 

This exact decomposition of $A$ is made possible by the fact that the four matrices $I$, $X$, $Y$, and $Z$ form a basis for the real vector space of $2\times 2$ Hermitian matrices. The real vector space of $2^n\times 2^n$ Hermitian matrices comprises the tensor product of $n$ instances of the $2\times 2$ matrix space, for which the set of matrices of the form $A_1\otimes\cdots\otimes A_n$---where $A_i\in\braces{I,X,Y,Z}$ for every $i$---forms a basis.

In general, the number of terms comprising $A$ is $O\paren{4^n}$; we shall consider observables for which the number of terms is $O(n^k)$ for some (reasonably small) $k$. We can impose the same constraint on the number of nonzero entries in $b$. Any product of local Pauli operators \begin{equation} c\paren{A_{1}\otimes\cdots\otimes A_{n}}\end{equation} is represented by a $2^n\times 2^n$ sparse matrix with exactly one nonzero entry in each row; this fact follows by induction on the number of factors $n$ in $c\paren{A_{1}\otimes\cdots\otimes A_{n}}$, using the definition of the Kronecker product in (\ref{kronecker}) alongside the fact that each of the $2\times 2$ Pauli matrices---and the identity matrix---has exactly one nonzero entry in each of its rows.

Given the index $x$ of some row of $c\paren{A_{1}\otimes\cdots\otimes A_{n}}$, it is possible to find the column index and value of the corresponding nonzero entry in $O(n)$ time. The heart of this idea may be found in appendix B of \cite{Choo_2020}, and in equation 16 of \cite{zhao2022quantumchemistry}. If $A$ comprises $O(n^k)$ terms in this local Pauli product form, then for each row of $A$ we may find the column indexes and values of all its nonzero entries in $O(n^{k+1})$ time by simply looking at each term individually. Since $(A\psi_\theta)(x)$ equals the inner product of row $x$ of $A$ with $\psi_\theta$, placing this constraint on $A$ ensures that we can always calculate the local energy of $A$ at any row in $O(n^{k+1})$ time.

\end{document}